\documentclass[journal,10pt]{IEEEtran}
\usepackage{cite}
\usepackage[pdftex]{graphicx}
\usepackage{amsmath}
\usepackage{algorithmic}
\usepackage{array}
\usepackage{url}
\usepackage{orcidlink}
\usepackage{hyperref}
\usepackage{makecell}
\usepackage[english]{babel} 
\usepackage{blindtext}
\usepackage{amsfonts}
\newcommand{\tDCF}{t\text{-}DCF}
\newcommand{\aDCF}{a\text{-}DCF}
\usepackage{booktabs}
\usepackage{xcolor}
\usepackage{multirow} 
\newcommand{\sep}[1][]{\,/\,}
\begin{document}
\title{Tandem spoofing-robust automatic speaker verification based on time-domain embeddings}

\author{Avishai~Weizman~\raisebox{1ex}{\orcidlink{0009-0004-1182-8601}},
        Yehuda~Ben-Shimol~\raisebox{1ex}{\orcidlink{0000-0002-4905-2085}} ,
        and Itshak~Lapidot~\raisebox{1ex}{\orcidlink{0000-0003-1066-3441}}}

\maketitle
\begin{abstract}
Spoofing-robust automatic speaker verification (SASV) systems are a crucial technology for the protection against spoofed speech.  In this study, we focus on logical access attacks and introduce a novel approach to SASV tasks.  A novel representation of genuine and spoofed speech is employed, based on the probability mass function (PMF) of waveform amplitudes in the time domain. This methodology generates novel time embeddings derived from the PMF of selected groups within the training set. This paper highlights the role of gender segregation and its positive impact on performance. We propose a countermeasure (CM) system that employs time-domain embeddings derived from the PMF of spoofed and genuine speech, as well as gender recognition based on male and female time-based embeddings. The method exhibits notable gender recognition capabilities, with mismatch rates of 0.94\% and 1.79\% for males and females, respectively. The male and female CM systems achieve an equal error rate (EER) of 8.67\% and 10.12\%, respectively. By integrating this approach with traditional speaker verification systems, we demonstrate improved generalization ability and tandem detection cost function evaluation using the ASVspoof2019 challenge database. Furthermore, we investigate the impact of fusing the time embedding approach with traditional CM and illustrate how this fusion enhances generalization in SASV architectures.
\end{abstract}

\renewcommand{\arraystretch}{1.2}
\section{Introduction} \label{sec:Introduction}
\IEEEPARstart{A} 
biometric system attempts to authenticate an individual's identity based on their behavioral and/or biological characteristics \cite{jain201650,hadid2015biometrics}. The characteristics of biometric recognition are categorized into two main types: anatomical and behavioral \cite{jain2006biometrics}. The anatomical features include face \cite{jain2011handbook}, fingerprint \cite{maltoni2009handbook}, palm print \cite{kong2009survey}, hand geometry \cite{sanchez2000biometric}, and ear shape \cite{yan2007biometric}, while gait \cite{yazdanpanah2010multimodal} and signature \cite{nalwa1997automatic} are some of the behavioral characteristics \cite{tolosana2015increasing}.
The evaluation of voice biometrics can be conducted through the examination of both anatomical and behavioral characteristics \cite{jain2006biometrics}. The primary goal of an \textit{automatic speaker verification} (ASV) system is to determine the veracity of an identity claim, whether it is true (target speaker) or false (non-target speaker). 
ASV systems provide users with a natural and non-invasive method of identity authentication, are widely used as a biometric identification tool, and are valued for their enhanced reliability, user-friendly interface, and ease of implementation. The common use of ASV systems in various telephony and communications networks makes voice vulnerable to spoofing attacks. 

Typically, spoofing involves an attempt to impersonate the target speaker and gain access to the system.
Spoofed speech samples can, mainly, be generated by methods such as speech synthesis, voice conversion, or playback of recorded speech \cite{kamble2020advances}.
There are two basic types of voice spoofing attacks: \textit{physical access} (PA) and \textit{logical access} (LA). PA attacks occur at the sensor (microphone) stage, while LA attacks attempt to attack the system after the sensor, at the feature representation stage, or at the modeling stage \cite{chadha2021review}.
Common types of LA attacks include \textit{speech synthesis} (SS), \textit{voice conversion} (VC) and their combinations. Speech synthesis takes text input and generates speech as output, while VC converts the voice of a source speaker to sound similar to the voice of a target speaker, as described in \cite{dutoit1997high,mohammadi2017overview}. 
Dedicated systems known as countermeasure systems have been developed to accomplish the task of detecting speech spoofing.
Recognizing the critical need for effective countermeasures, various anti-spoofing task challenges have been established over the years.
Several challenges have been conducted with CM systems or SASV systems.
where the goal is to distinguish between real and spoofed utterances produced by SS and VC algorithms.

This work emphasizes LA attacks, utilizing the widely used ASVspoof2019 LA database as its primary benchmark for evaluation \cite{wang2020asvspoof}.
The majority of features utilized in CM systems are frequency-based, such as LFCC \cite{mohammadi2017robust,kamble2020advances,mittal2022automatic}, CQCC \cite{todisco2017constant,kamble2020advances,mittal2022automatic}, spectrogram \cite{xie2024temporal,hernandez2023voice,boyd2023voice}, and Mel-spectrogram \cite{rohdin2024but,boyd2023voice}. In \cite{lavrentyeva2019stc,karo2023compact}, the PMF-based time-domain feature extraction method was presented.
Recently, models based \textit{self supervised learning} (SSL), such as Wav2Vec and HuBERT, have been increasingly applied in countermeasure systems due to their ability to learn robust representations from raw speech data \cite{baevski2020wav2vec,xie2024temporal,combei2024wavlm,rohdin2024but}. Fusion methods are also commonly used to improve performance and generalization, and in some cases, both SSL-based models and fusion techniques are combined to further enhance system effectiveness \cite{rohdin2024but,xie2024temporal,xu2024szu}.
On the classification side, advances include the development of novel loss functions, as well as loss functions specifically tailored from computer vision techniques for this task. Two such examples are \textit{one-class softmax} {(OCS) loss \cite{zhang2021one}} and \textit{AMSoftmax} (AMS) loss \cite{wang2018additive}.

We have found out that gender based SASV can be beneficial. It is reasonable to assume that the training set for anti-spoofing and verification tasks will include prior knowledge of gender identity. Furthermore, in several scenarios, this information can also be available during the verification task. The process of identifying the gender of the input individual is referred to as a gender classification or recognition task. By leveraging gender information, CM and ASV systems can develop for each gender separately and adjust the threshold for each specific gender, thereby enhancing overall performance. In recent years, there has been a growing interest in the gender recognition task, leading to numerous studies being conducted, as noted in \cite{yucesoy2024gender}. Previous research has primarily focused on identifying optimal features and designing effective gender recognition models. Commonly used features include MFCC, Mel-spectrogram, pitch, and other voice parameters \cite{chachadi2022voice,raahul2017voice,yucesoy2024gender}. This study will demonstrate that time-domain based embeddings are also effective for gender recognition tasks.

In \cite{karo2023compact, weizmanCSCML}, the authors devised a time-domain embedding methodology to differentiate between authentic and imitated speech by utilizing the PMF of filtered signals. These studies concentrated on the time embedding generation process and the deployment of a basic \textit{logistic regression} (LR) model for the classifier in the CM system.

The primary objective of our research is to develop a novel SASV system that integrates data from CM-based time embeddings and conventional ASV systems. Our study introduces an SASV system that employs time-domain embeddings and illustrates the value of a gender-specific CM approach. To the best of our knowledge, this research represents the first effort to construct a SASV system based on time embeddings. The following step is to combine time-based embeddings with frequency-based embeddings. This will allow the system to learn complementary information from both domains and improve its ability to generalize to unseen spoofing attacks. {Finally, we demonstrate that integrating this time-domain approach with existing frequency-based CM systems.}

The remainder of the paper is organized as follows: \autoref{sec:Time_embeedings_based_PMF} describes the PMF-based time embeddings. 
\autoref{sec:SASV_description} describes the SASV system, and details its components.
In \autoref{sec:Database_and_evaluation_metrics}, we describe the LA database from the ASVspoof2019 challenge that was used in this work, along with the evaluation metrics used for performance assessment.
The experiments and results for the SASV system and its components are presented in \autoref{sec:Experiments_and_results}. 
Finally, \autoref{sec:Conclusions_and_future_work} summarizes our findings, presents our conclusions, and discusses potential directions for future research.

\section{Time embeddings based PMF} \label{sec:Time_embeedings_based_PMF}
The fundamental concept of time embedding is to utilize the dissimilarities in amplitude distributions between authentic and spoofed speech \cite{lapidot2019effects,karo2023compact,karo24_odyssey,weizmanCSCML} (or other groups within the database \cite{lapidot2018speech}) to categorize signals based on the degree of similarity between the input signal and the aforementioned speech groups.

The PMF is approximated at $x_0$ by 
\begin{equation} \label{eq:approx_PMF}
    \mathrm{PMF}(x=x_{0}) = \int_{x_{0}-\frac{\Delta x}{2}}^{x_{0}+\frac{\Delta x}{2}} \mathrm{PDF}(\theta)\, d\theta
\end{equation}
For waveform samples with $b$ bits per sample, there are $2^b$ possible discrete values. In ASVspoof2019, LA audio files were sampled with $16$ bits per sample and subsequently normalized to the $[-1,1]$ range, which defines the histogram bins for the calculation of the PMF. Following \cite{karo2023compact,karo24_odyssey}, for filtered data we clip the data to the range $[-1,1]$, and use $2^{16}$ bins.

\begin{figure}[t]
    \centering
    \includegraphics[width=\linewidth]{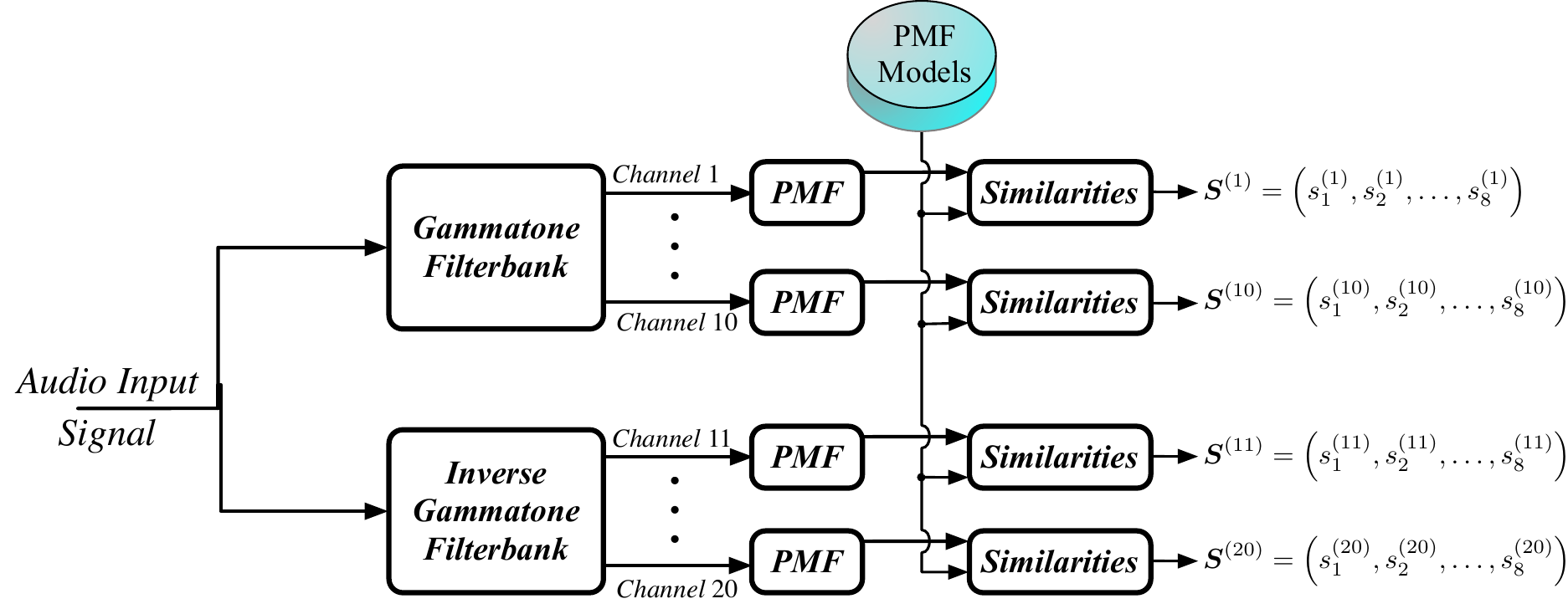}
    \caption{Time embedding generation process \cite{weizmanCSCML}.}
\label{fig:Features_extraction_scheme_matan}
\end{figure}

The PMF models are derived from the statistical data obtained from the training set. The PMF models are calculated for all files within a predefined group. For example, a predefined group may be based on gender or the distinction between spoofed and genuine speech. An additional example is the generation of models comprising more than two categories: male spoofed, female spoofed, genuine male, and genuine female speech.
In this work, we followed the approach in \cite{karo2023compact,karo24_odyssey}.
Each speech signal is subjected to processing through $10$ Gammatone and $10$ Inverse Gammatone filter banks (gathered into $N=20$ filters), and the PMF of the output of each filter is calculated. Subsequently, similarity measures are calculated between the PMF of the new signal and the class PMF models. Our approach encompasses eight distinct similarity measures, including the Quadratic Chi-distance \cite{pele2010quadratic}, Normalized Cross Correlation \cite{briechle2001template}, Hellinger Distance \cite{hellinger1909neue}, Intersection, Kullback-Leibler divergence, \textit{symmetrized Kullback-Leibler divergence} (KLS), \textit{Jensen-Shannon divergence} (JSD), and modified Kolmogorov-Smirnov divergence, as described in \cite{karo2023compact}. The PMFs are expressed as:
\begin{equation*}
  \{\mathrm{PMF}^{(n)}_X: X \in \{ \mathrm{input}, \mathrm{class_1}, \mathrm{class_2} \}, n = 1,\ldots,N\}
\end{equation*}
where $\mathrm{class_1}$ and $\mathrm{class_2}$ can be \textit{genuine} and \textit{spoofed} or \textit{male} and \textit{female}. For each trial, the similarity is computed for each PMF model.
\begin{equation*}
d_{l}\left(\mathrm{PMF}^{(n)}_{\mathrm{input}},\mathrm{PMF}^{(n)}_{\mathrm{class_q}}\right),\ \  q\in\{1, 2\}
\end{equation*}
and $l$ represents the specific type of similarity measure from the eight previously described. Next, an embedding vector is generated for each audio recording:
\begin{multline}
\label{eq:embeddingsVector}
  s^{(n)}_{l} = d_{l}\left(\mathrm{PMF}^{(n)}_{\mathrm{input}}, \mathrm{PMF}^{(n)}_{\mathrm{class_2}}\right) - \\ d_{l}\left(\mathrm{PMF}^{(n)}_{\mathrm{input}}, \mathrm{PMF}^{(n)}_{\mathrm{class_1}}\right)
\end{multline}
The outcome of each trial is an embedding vector of size 160, derived from the outputs of 10 Gammatone and 10 Inverse Gammatone filters, with 8 similarity measures per PMF (i.e., 20 channels $\times$ 8 measures). \autoref{fig:Features_extraction_scheme_matan} depicts the process of our time embedding generation. {Note that the PMF-based time embedding generation approach is also used for gender recognition, as shown in \autoref{fig:Anti_spoofing_System_arc}-A.}

\section{Spoofing-robust automatic speaker verification system description}  \label{sec:SASV_description}
\begin{figure*}[t]
    \centering
   \includegraphics[width=0.7\textwidth]{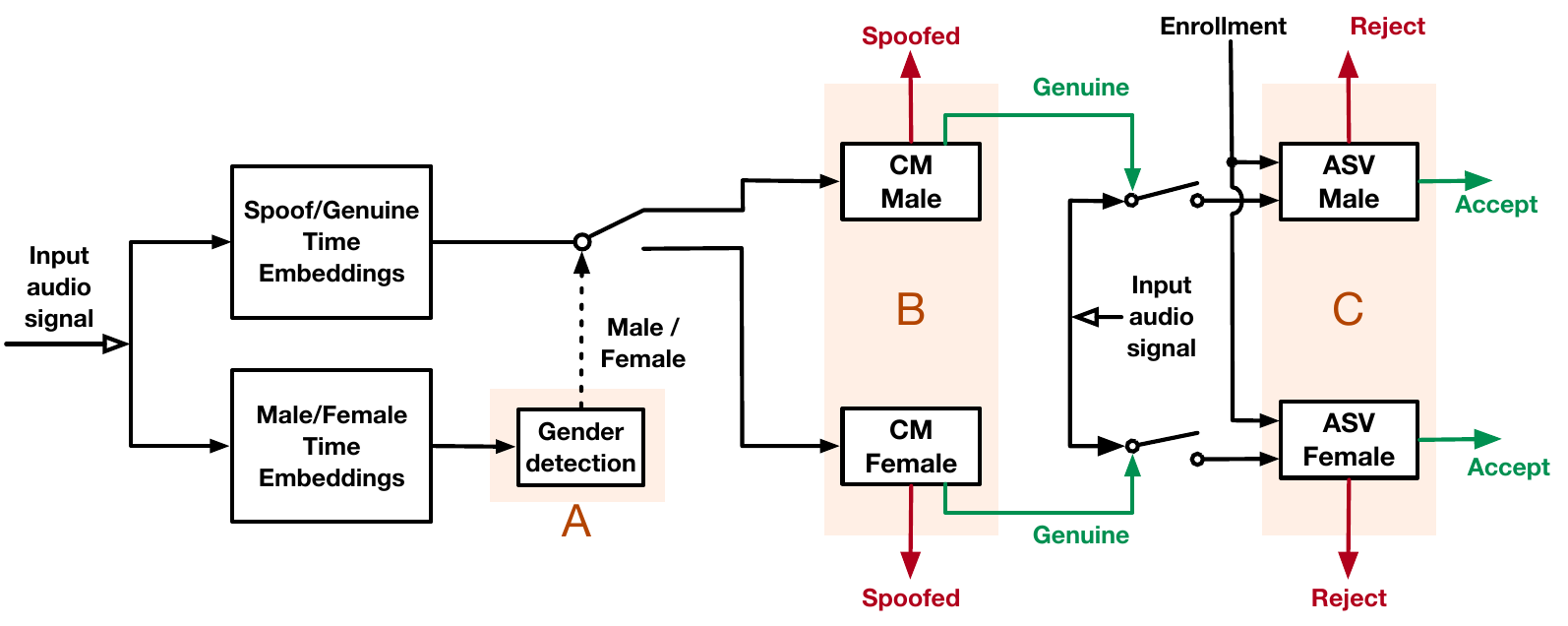}
    \caption{The Spoofing-robust speaker verification architecture. The same audio signal, is fed into the time embedding generation system (on the left) and the ASV system (on the right).} 
    \label{fig:Anti_spoofing_System_arc}
\end{figure*}
The proposed SASV system is depicted in \autoref{fig:Anti_spoofing_System_arc} and is comprised of three sub-systems. The first sub-system (A) classifies the input speech signal as male or female gender, based on the findings of \cite{karo2023compact}, which demonstrated that the PMFs vary by speaker gender. The second subsystem (B) applies the CM to each gender class. Sub-systems A and B use the time-domain embeddings as their input. Subsystem C is the ASV system, based on the ECAPA-TDNN architecture, which accepts both the target speaker enrollment data and the input audio signal of the claimed speaker. The following subsections describe the proposed SASV subsystems.
\subsection{Gender Recognition System}
Based on the findings in \cite{karo2023compact}, a gender recognition system is beneficial to leverage the differences between spoofed and genuine speech for each gender separately.
The proposed gender recognition system directs input to the relevant gender-based module, employing time embeddings derived from gender group models. The gender recognition system was implemented with the \textit{extreme gradient boosting} (XGB) classifier \cite{chen2015xgboost}.
This classifier demonstrated the best performance on the ASVSpoof2019 database among the tested classifiers, which included: \textit{logistic regression} (LR), \textit{k-nearest neighbors} (kNN), \textit{random forest} (RF), \textit{support vector machine} (SVM), balanced RF, XGB, and \textit{random under-sampling AdaBoost algorithm} (RUSBoost). Each classifier was trained on the training set and tuned using grid search on the development set.
\subsection{Countermeasures Systems}
\begin{figure}[t]
\centering
\includegraphics[width=8.6cm]{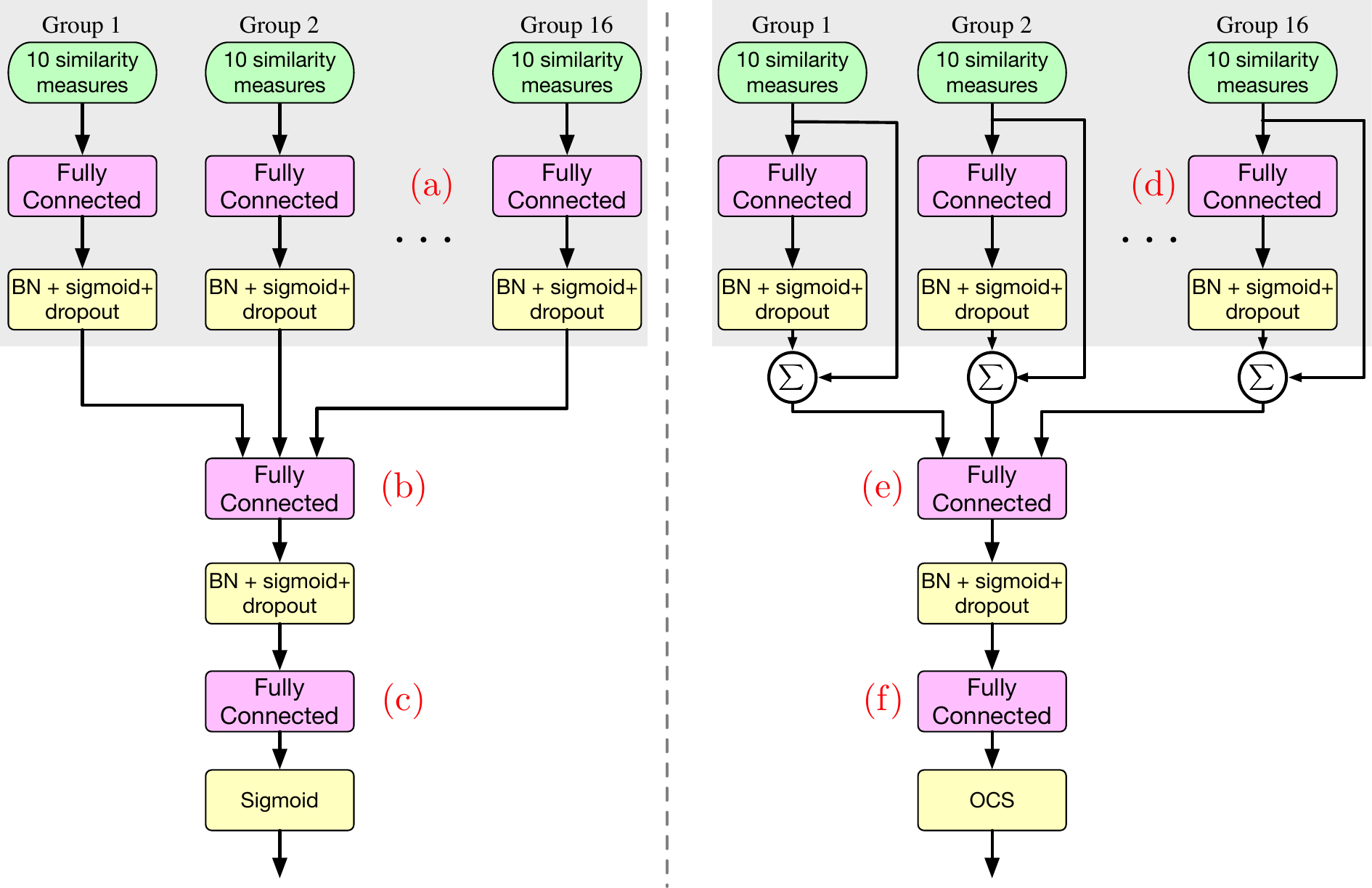}
\caption{Neural networks structures: left side for the male gender and the right for the female\sep[]both genders.}
\label{fig:Female_Male_CM}
\end{figure}

\autoref{fig:Female_Male_CM} illustrates the proposed gender dependet (GD) CM models based on time embeddings. The network is not overly complex due to the extensive pre-processing required to generate the time embeddings. Subsequently, the generated time embedding is partitioned into 16 groups based on the similarity metrics, as described in~\autoref{sec:Time_embeedings_based_PMF}. The distinction between Gammatone filters and inverse Gammatone filters is maintained. The objective of this approach is to prevent over fitting to specific attack types and to enhance the ability to generalize to new attacks in the evaluation set. This is achieved by separating the different similarity metrics and feeding each group into a separate neural network. 
The CM for the male gender was implemented using fully connected (FC) layers and a sigmoid activation function. For the female gender and gender-independent (GI), a configuration comprising fully connected layers with residual connections and a one-class softmax loss function was identified as superior \cite{zhang2021one}. The incorporation of residual connections with FC layers was informed by \cite{orhan2017skip}. In this configuration, the CM sub-system serves as a gate to the ASV system for each gender class. These configurations were identified following an extensive examination of numerous alternative configurations, including activations, regularizations, losses, and other factors.
\subsection{Automatic Speaker Verification Systems}
The objective of ASV systems is to ascertain whether the claimed speaker is a target speaker or an impostor. The ECAPA-TDNN architecture from the SpeechBrain framework \cite{speechbrain} was employed as the basis for our ASV systems, which were trained on the VoxCeleb1 and VoxCeleb2 databases \cite{desplanques2020ecapa}. The aforementioned gender-separated ASV systems afford complete flexibility during experimentation on the ASVSpoof2019 database. This flexibility enables the threshold to be matched for each gender individually. The speaker verification task was accomplished by calculating the cosine distance between speaker embeddings. In instances where multiple embeddings were available during the enrollment stage, the embeddings were averaged to form the speaker's ID registration. The ASV systems' detection thresholds were fixed according to the EER threshold for each gender on the evaluation set, as described in \cite{yamagishi2019asvspoof}. It is crucial to note that ECAPA-TDNN was not gender-adapted; only the threshold was set per gender.
\section{Database and evaluation metrics} \label{sec:Database_and_evaluation_metrics}
This section overviews of the ASVSpoof2019 database and the evaluation metrics commonly used in anti-spoofing and speaker verification tasks. These include the \textit{detection cost function} (DCF), \textit{tandem DCF} (t-DCF), and \textit{agnostic DCF} (a-DCF), along with the confidence intervals method.
\subsection{ASVSpoof2019 LA Database}
The present study is focused on the LA attacks from the ASVSpoof2019 dataset \cite{wang2020asvspoof}.
The LA database is divided into three distinct datasets, namely, training, development, and evaluation. The ASVspoof2019 database comprises speech data from 107 speakers, of whom 46 were males, and 61 were females. The training set comprises data from 20 speakers (8 males, 12 females), the development set from 20 speakers (8 males, 12 females), and the evaluation set from 67 speakers (30 males, 37 females).
The ASVspoof2019 LA dataset was constructed with a diverse array of 17 TTS and VC attacks, as well as their combinations. Of these, 11 attacks are unknown spoofing attacks in the evaluation set, while six are known attacks included in the training and development sets. Furthermore, two of the unknown attacks in the evaluation set employ the same algorithms as two of the known attacks from the training set but differ in their data sources.

\subsection{Evaluation Metrics}
The most commonly utilized metric for evaluating biometrics research is the EER, which serves as an upper bound to the Bayes error rate \cite{kinnunen2023t,brummer2021out}. This metric was employed in all ASVSpoof challenges. Another frequently utilized metric for the assessment of speaker verification systems is the DCF, which is employed to provide a balanced performance measure that considers both false alarms and missed detections. The third evaluation metric, the t-DCF, and its variants were initially introduced in the ASVspoof2019 challenge \cite{yamagishi2019asvspoof,kinnunen2020tandem}. The last metric is the a-DCF, which reflects the cost of decisions in a Bayes' risk sense, similar to the DCF, with explicitly defined class priors and detection cost model \cite{shim24_odyssey}.

\subsubsection*{DCF} DCF is Bayesian risk classification for a given score's threshold $\theta$. It is broadly used for evaluating ASV systems according to their minimum DCF values \cite{doddington2000nist}. The DCF is defined by
\begin{equation}
    \label{eq:DCF}
    \mathrm{DCF}(\theta) = \mathrm{C}^{\mathrm{asv}}_{\mathrm{miss}} \pi_{\mathrm{tar}} \mathrm{P}^{\mathrm{asv}}_{\mathrm{miss}}\left(\theta\right) + \mathrm{C}^{\mathrm{asv}}_{\mathrm{fa}} (1 - \pi_{\mathrm{tar}}) \mathrm{P}^{\mathrm{asv}}_{\mathrm{fa}}\left(\theta\right)
\end{equation}
where $\mathrm{C}^{\mathrm{ASV}}_{\mathrm{miss}}$ and $\mathrm{C}^{\mathrm{ASV}}_{\mathrm{fa}}$ denote the costs for a \textit{miss} and \textit{false alarm} errors, respectively. $\pi_{tar}$ represents the prior probability of the \textit{target}, while $\mathrm{P}^{\mathrm{ASV}}_{\mathrm{miss}}\left(\theta\right)$ and $\mathrm{P}^{\mathrm{ASV}}_{\mathrm{fa}}\left(\theta\right)$ are the empirical probabilities of \textit{missed} and \textit{false alarm} events for a given threshold $\theta$.

\subsubsection*{t-DCF Unconstrained}The t-DCF is an evaluation metric used to assess the performance of an integrated tandem system comprising a CM and an ASV system  \cite{kinnunen2020tandem}. In case it is feasible to modify both thresholds of the CM  and ASV, namely, ($\tau_{\mathrm{cm}}$) and ($\tau_{\mathrm{asv}}$), respectively, the SASV system can be adjusted to achieve optimal compatibility. The probabilities associated with the unconstrained t-DCF expression are defined by:
\begin{equation}
\label{eq:probabilities}
    \begin{aligned}
    &\mathrm{P_a}(\tau_{\mathrm{cm}},\tau_{\mathrm{asv}}) = (1 - \mathrm{P_{\mathrm{miss}}^{\mathrm{cm}}}(\tau_{\mathrm{cm}}))\cdot \mathrm{P_{\mathrm{miss}}^{\mathrm{asv}}}(\tau_{\mathrm{asv}}) \\ 
    &\mathrm{P_b}(\tau_{\mathrm{cm}},\tau_{\mathrm{asv}}) = (1 - \mathrm{P_{\mathrm{miss}}^{\mathrm{cm}}}(\tau_{\mathrm{cm}}))\cdot \mathrm{P_{\mathrm{fa}}^{\mathrm{asv}}}(\tau_{\mathrm{asv}}) \\ 
    &\mathrm{P_\mathrm{c}}(\tau_{\mathrm{cm}},\tau_{\mathrm{asv}}) = \mathrm{P_{\mathrm{fa}}^{\mathrm{cm}}}(\tau_{\mathrm{cm}})\cdot \mathrm{P_{\mathrm{fa,spoof}}^{\mathrm{asv}}}(\tau_{\mathrm{asv}}) \\ 
    &\mathrm{P_\mathrm{d}}(\tau_{\mathrm{cm}},\tau_{\mathrm{asv}}) =  \mathrm{P_{\mathrm{miss}}^{\mathrm{cm}}}(\tau_{\mathrm{cm}})
\end{aligned}
\end{equation}
where each probability corresponds to a particular error type.
The combination of class priors, costs, and error terms yields the unconstrained t-DCF:
\begin{equation}
\label{eq:unconstrained}
    \begin{aligned}
    &\mathrm{\tDCF}(\tau_\mathrm{cm},\tau_\mathrm{asv}) = 
    \mathrm{C_{miss}}\pi_\mathrm{tar} \left[\mathrm{P_a}(\tau_\mathrm{cm}, \tau_\mathrm{asv}) +\mathrm{P_d}(\tau_\mathrm{cm},\tau_\mathrm{asv})\right]\\
    &+\mathrm{C_{fa}}\pi_\mathrm{non} \mathrm{P_b}(\tau_\mathrm{cm},\tau_\mathrm{asv})+\mathrm{C_{fa,spoof}}\pi_\mathrm{spoof}\mathrm{P_c}(\tau_\mathrm{cm},\tau_\mathrm{asv})
    \end{aligned}
\end{equation}
where the three elements in \eqref{eq:unconstrained} correspond to target, non-target, and spoof-related errors, respectively.
The normalized t-DCF unconstrained case is defined by
\begin{equation}
    \label{eq:norm_t_dcf_uncon}
    \begin{aligned}
        \mathrm{\tDCF_{normalize}^{Unconst}}(\tau_{\mathrm{cm}}, \tau_{\mathrm{asv}}) &= 
        \frac{\mathrm{\tDCF}^{\mathrm{Unconst}}(\tau_{\mathrm{cm}}, \tau_{\mathrm{asv}})}
        {\mathrm{\tDCF}_{\mathrm{default}}^{\mathrm{Unconst}}(\tau_{\mathrm{cm}}, \tau_{\mathrm{asv}})}
    \end{aligned}
\end{equation}
where $\mathrm{\tDCF_{\mathrm{default}}^{\mathrm{Unconst}}} > 0 $ is the t-DCF of a default (reference) system. The default system is selected based on the system parameters and the $\mathrm{\tDCF_{\mathrm{default}}^{Unconst}}$ is defined by
\begin{multline} \label{eq:unc_default}
      \mathrm{\tDCF_{\mathrm{default}}^{Unconst}}(\tau_{\mathrm{cm}}, \tau_{\mathrm{asv}}) = \\
        \min(\mathrm{C_{\mathrm{fa}}}\pi_{\mathrm{non}} 
        + \mathrm{C_{\mathrm{fa,spoof}}}\pi_{\mathrm{spoof}}, 
        \mathrm{C_{\mathrm{miss}}}\pi_{\mathrm{tar}})
\end{multline}

\subsubsection*{t-DCF ASV-Constrained} Following \cite{kinnunen2018t}, the CM threshold is adjustable, whereas the ASV threshold is fixed at the EER operating point, which also serves as the upper bound to the Bayes error rate \cite{yamagishi2019asvspoof}.
\begin{equation}
\label{eq:constrained}
    \mathrm{\tDCF}(\tau_\mathrm{cm}) = \mathrm{C_0}+\mathrm{C_1}\cdot \mathrm{P_{miss}^{cm}}(\tau_{cm}) + \mathrm{C_2\cdot P_{fa}^{cm}}(\tau_{cm})
\end{equation}
The constants $\mathrm{C_0,C_1,C_2}$, are determined by both the t-DCF's parameters and the ASV error rates, as demonstrated by
\begin{equation}
\centering
\label{eq:constrained_prop}
\begin{aligned}
    \mathrm{C_0} &= \pi_{\mathrm{tar}}\mathrm{C_{miss}}\mathrm{P^{asv}_{miss}(\tau_{asv})} + \pi_{\mathrm{nontar}}\mathrm{C_{fa}}\mathrm{P^{asv}_{fa}(\tau_{asv})} \\
    \mathrm{C_1} &= \pi_{\mathrm{tar}}\mathrm{C_{miss}} - \\
     & \hphantom{XX} \left(\pi_{\mathrm{tar}}\mathrm{C_{miss}}\mathrm{P^{asv}_{miss}}(\tau_\mathrm{asv}) - \pi_{\mathrm{nontar}}\mathrm{C_{fa}}\mathrm{P^{asv}_{fa}(\tau_{asv})}\right) \\
    \mathrm{C_2} &= \pi_{\mathrm{spoof}}\mathrm{C_{fa,spoof}}\mathrm{P^{asv}_{fa,spoof}}
\end{aligned}
\end{equation}
$\mathrm{C_0}$ represents the operating point of \eqref{eq:constrained}, and is solely dependent on the ASV threshold, which is fixed in the ASV-constrained t-DCF expression.
The normalized version of the t-DCF for the ASV-constrained case is defined by
\begin{equation} 
\label{eq:t_dcf_asv_const}
\mathrm{\tDCF_{\mathrm{normalize}}^{\mathrm{Constr}}}(\tau_{\mathrm{cm}}) = \frac{\mathrm{\tDCF^{\mathrm{Constr}}}(\tau_{\mathrm{cm}})}{\mathrm{\tDCF_{\mathrm{default}}^{\mathrm{Constr}}}(\tau_{\mathrm{cm}})}
\end{equation}
and the default system is selected based on the system parameters as given by
\begin{equation} \label{eq:def_t_dcf_asv_const}
\mathrm{\tDCF_{\mathrm{default}}^{Constr}}(\tau_{\mathrm{cm}}) = \mathrm{C_{0}} + \mathrm{\min(C_{1}, C_{2}}) 
\end{equation}

\subsubsection*{a-DCF} 
The a-DCF metric is a performance measure introduced for SASV systems. In contrast to the t-DCF, the a-DCF is not constrained by the system's structural configuration and thus is not confined to the tandem structure \cite{shim24_odyssey,delgado2024asvspoof}.  In contrast to the t-DCF's ineffectiveness in an integrated system, the a-DCF can be applied to any system configuration, including a tandem system. As stated in \cite{shim24_odyssey}, in a tandem configuration, the scores are generated based on the CM threshold, as shown by:
\begin{equation}
    \begin{aligned}
        \mathrm{s} = 
        \begin{cases} 
        \mathrm{s}_{\mathrm{asv}}, & \mathrm{if }\ \mathrm{s}_{\mathrm{cm}} \geq \tau_{\mathrm{cm}}, \\
        -\infty, & \mathrm{if\ } \mathrm{s}_{\mathrm{cm}} < \tau_{\mathrm{cm}}.
        \end{cases}
    \end{aligned}
\label{eq:a_dcf_tandem}
\end{equation}
Here, $\mathrm{s_{asv}}$ and $\mathrm{s_{cm}}$ are the detection scores produced by the speaker verification system (ASV sub-system) and the spoof detection system (CM sub-system), respectively, and $\mathrm{\tau_{cm}}$ is a fixed CM threshold.
The a-DCF is defined by:
\begin{multline}
        \mathrm{\aDCF}(\tau) = \mathrm{C}_\mathrm{miss} 
        \pi_\mathrm{tar} P_\mathrm{miss}(\tau)+ \\
        \mathrm{C}_\mathrm{fa,non} \pi_\mathrm{non} P_\mathrm{fa,non}(\tau) 
        + \mathrm{C}_\mathrm{fa,spf} \pi_\mathrm{spf} P_\mathrm{fa,spf}(\tau)
    \label{eq:a_DCF}
\end{multline}
where $\mathrm{C}_{\mathrm{miss}}$, $\mathrm{C}_{\mathrm{fa,non}}$, and $\mathrm{C}_{\mathrm{fa,spf}}$ are the costs of target speaker missing (falsely rejecting), falsely accepting a non-target speaker (zero effort error), and falsely accepting a spoof, respectively.
$\pi_\mathrm{tar}$, $\pi_{\mathrm{non}}$, and $\pi_{\mathrm{spf}}$ are the assumed class priors. $\mathrm{P}_{\mathrm{miss}}(\tau)$, $\mathrm{P}_{\mathrm{fa,non}}(\tau)$, and $\mathrm{P}_{\mathrm{fa,spf}}(\tau)$ represent the respective empirical detection error rates at the detection threshold $\tau$ of the ASV system.

\subsubsection*{Confidence intervals} 
The evaluation of machine learning models entails the calculation of a pertinent performance metric through the utilization of an evaluation database that represents the target scenario \cite{raschka2018model}. However, the metric may vary due to factors such as the training data, the evaluation data, and the random seeds, which could impact the conclusions drawn about system performance. To account for this variability, we employ bootstrapping, as outlined by \cite{Confidence_Intervals}, to compute confidence intervals. In this process, the evaluation set is passed through the system, and output scores and labels are sampled with replacement to generate new sets, forming an empirical distribution of the chosen metric values. This distribution is then used to calculate confidence intervals \cite{poh2007estimating}. This method circumvents assumptions about data distribution and offers a more flexible approach than traditional statistical methods \cite{keller2005benchmarking}. All results presented in this paper have been computed with $\alpha = 5$, which represents the confidence level. The confidence interval has been calculated in the range $[\alpha/2,100 - \alpha/2]$ percentiles, using $M = 1000$ bootstrapping iterations.

\section{Experiments and results} \label{sec:Experiments_and_results}
This section presents the experiments conducted and the results obtained in the course of this research.

\subsection{Gender Recognition based Time Embeddings}
\label{subsec:GenderRecognition}
\begin{figure}[t]
    \centering
    \includegraphics[width=8.6cm]{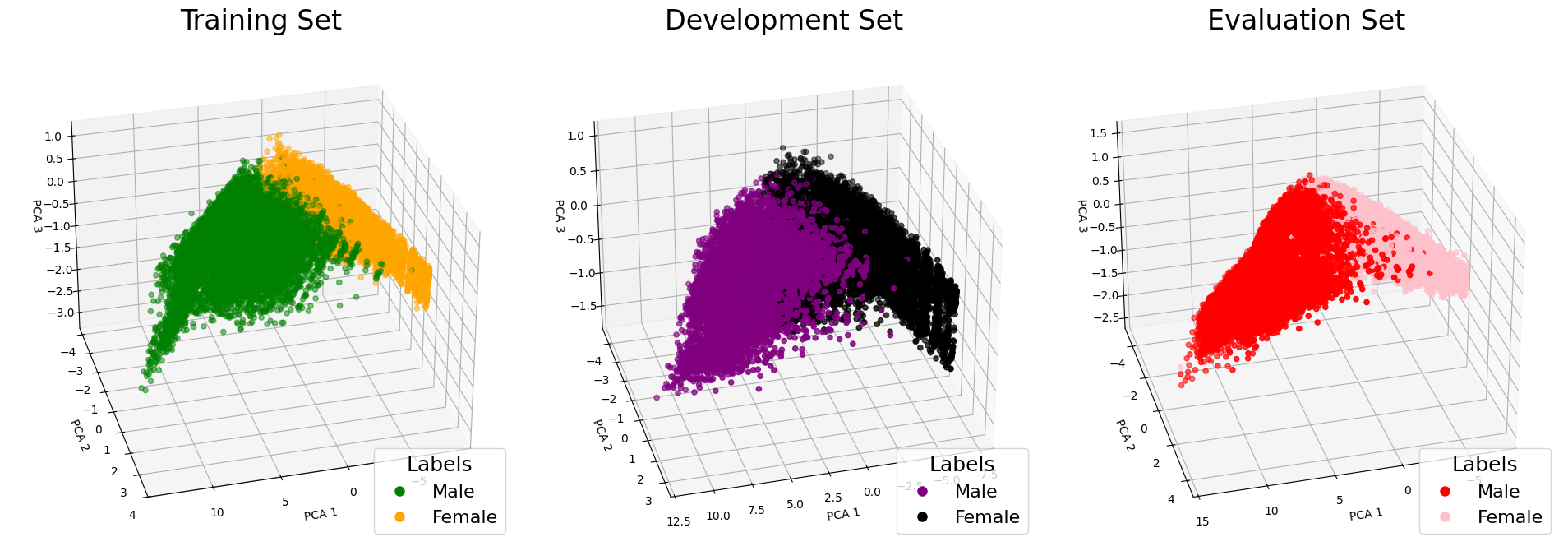}
    \caption{3 dimensional PCA projection  of time embeddings grouped by gender \cite{weizmanCSCML}.}
    \label{fig:Gender_Time_Embeddings_PCA}
\end{figure}

The use of gender-based segregation enables the exploitation of the distinctive characteristics of spoofed and genuine speech for each gender on an individual basis \cite{karo2023compact}. In the PMF-based time embedding approach, models in the database can be grouped based on shared characteristics of speech signals, such as gender. In this context, model groups represent male and female categories using a PMF-based time embedding approach, thereby enabling the calculation of similarity measures between a claimed speech signal and each gender class. The visualization of these gender time embeddings, which were reduced to $3$ dimensions space using the \textit{Principal Component Analysis} (PCA), \cite{abdi2010principal}, is depicted in \autoref{fig:Gender_Time_Embeddings_PCA}. \autoref{fig:Gender_Time_Embeddings_PCA} illustrates the separation between males and females based on time embedding from gender model groups across different sets.
On all three subsets (i.e., training, development, and evaluation), a clear separation can be observed in the $3$ dimensions space.
The accuracy of several classifiers for gender classification based on time-domain embeddings was evaluated, with the results presented in \autoref{tab:Male_vs_Female}. 
\begin{table}[t]
    \centering
    \caption{Comparison between gender recognition classifiers applied to the eval set.}
    \label{tab:Male_vs_Female}
    \begin{tabular}{ccc}
        \hline
        \multirow{ 2}{*}{}\textbf{Model} & \textbf{PMF based}  & \textbf{PMF based}  \\
         & \textbf{Spoofed\sep[]Genuine} & \textbf{ Male\sep[]Female}\\
        \hline
        \multirow{2}{*}{LR} & 0.94\%\sep[]3.97\% & 1.37\%\sep[]1.69\% \\ & 
        {\scriptsize [0.50,1.65] \sep[] [1.91,6.78]} & {\scriptsize [0.32,2.96] \sep[] [0.70,3.21]} \\
         \multirow{2}{*}{KNN} & 18.3\%\sep[]11.76\% & 4.11\%\sep[]2.50\% \\ & 
        {\scriptsize [16.16,20.61] \sep[] [7.29,16.72]} & {\scriptsize [2.72,5.93] \sep[] [1.00,4.67]} \\
        \multirow{2}{*}{SVM} & 1.08\%\sep[]3.28\% & 1.59\%\sep[]1.57\% \\ & 
        {\scriptsize [0.48,1.90] \sep[] [1.51,5.78]} & {\scriptsize [1.24,2.80] \sep[] [0.59,4.12]} \\
        \multirow{2}{*}{RF} & 2.35\%\sep[]3.01\% & 1.52\%\sep[]1.97\% \\ & 
        {\scriptsize [1.16,3.94] \sep[] [1.48,5.11]} & {\scriptsize [0.02,0.48] \sep[] [0.00,0.04]} \\ 
        \multirow{2}{*}{Balanced RF} & 1.59\%\sep[]3.67\% & 1.17\%\sep[]2.36\% \\ & 
        {\scriptsize [0.76,2.69] \sep[] [1.83,6.24]} & {\scriptsize [1.05,2.24] \sep[] [0.66,4.23]} \\
        \multirow{2}{*}{RUSBoost} & 1.03\%\sep[]3.05\% & 1.02\%\sep[]1.93\% \\ & 
        {\scriptsize [0.44,1.75] \sep[] [1.49,5.05]} & {\scriptsize [0.30,2.09] \sep[] [0.75,3.52]} \\
        \multirow{2}{*}{XGB} & 0.77\%\sep[]3.04\% & 0.94\%\sep[]1.79\% \\ & 
        {\scriptsize [0.37,1.29] \sep[] [1.46,5.06]} & {\scriptsize [0.37,1.68] \sep[] [0.64,3.29]} \\ 
        \hline
    \end{tabular}
\end{table}

Before extracting the time-domain gender embeddings, we tried to apply the time-domain spoof embeddings, since we already have them ``for free''. From~\autoref{tab:Male_vs_Female}, we see that spoof embeddings contain a lot of gender information and have high discriminative power. Nevertheless, gender embeddings show better classification accuracy for the gender recognition task compared to spoof embeddings.
Based on the results presented in~\autoref{tab:Male_vs_Female}, the XGB classifier was selected for the SASV system due to its superior performance, compared to all the other classifiers.

\subsection{Countermeasures based on Time Embeddings}
\begin{figure}[t]
\centering
 \includegraphics[width=8.6cm]{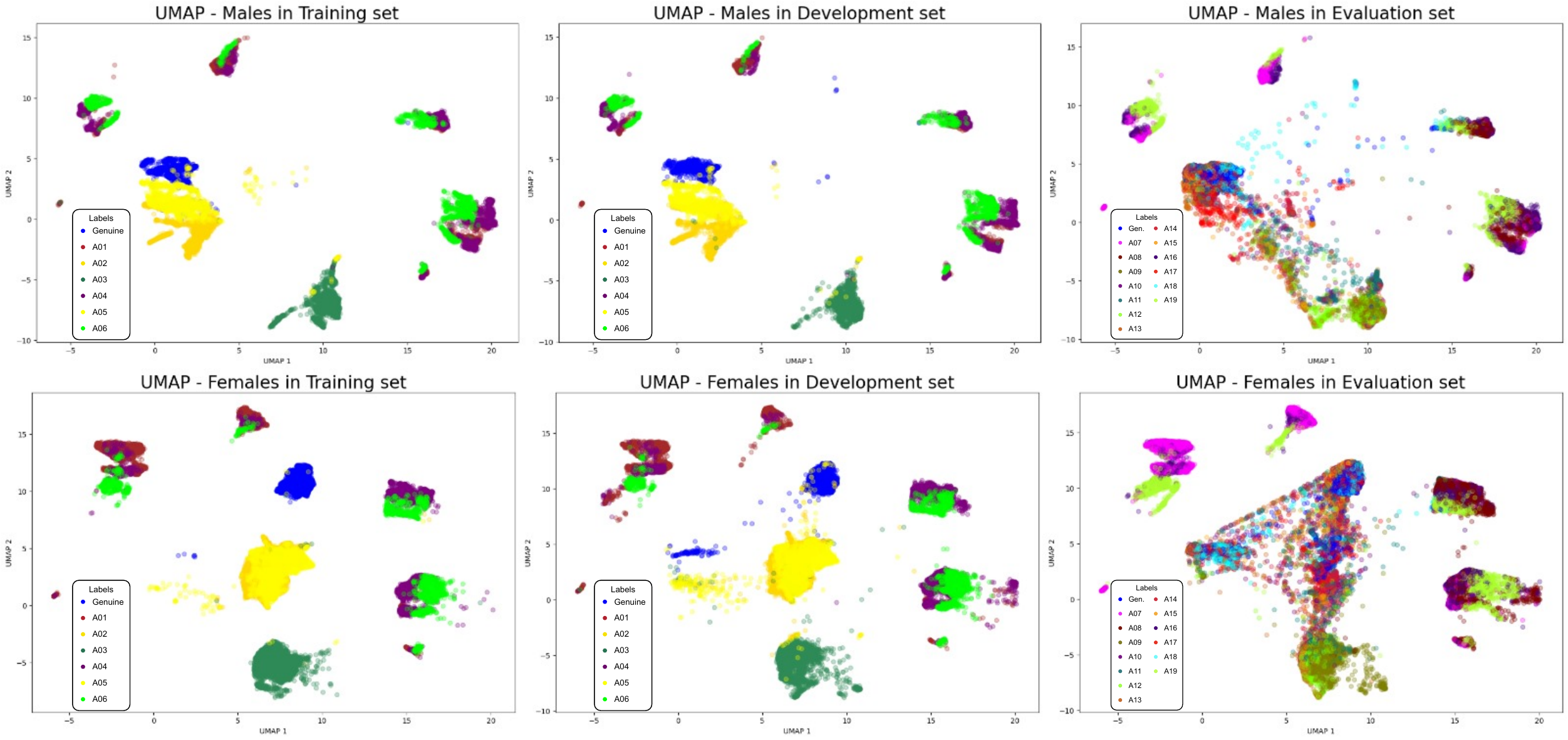} 
 \caption{UMAP dimensionality reduction of time embeddings trained on the training set, applied for each set, and visualized for each distinct gender group.}
\label{fig:umap_time_embedding}
\end{figure}
The visualization of the time embedding based on the PMF models of each attack was generated using the \textit{Uniform Manifold Approximation and Projection} (UMAP) algorithm 
\cite{mcinnes2018umap} to reduce the dimensions to 2.
The UMAP dimensionality reduction algorithm was performed with the parameters set to 15 nearest neighbors, a minimum distance of 0.1, and the Euclidean distance metric. The resulting two-dimensional space is shown in \autoref{fig:umap_time_embedding}.
The UMAP algorithm was trained on the training set and applied to each set, and the results are presented for males and females. Genuine speech is represented in a blue color in all the plots.
The separation between genuine and spoofed speech is clear in the training and development sets, in contrast to the lack of clear separation in the evaluation set.
In this 2D space, mismatch attacks differ between the training and development sets compared to the evaluation set, while genuine speech remains in the same location.
This result is expected, since the models were trained and validated on the same set of attacks (A01-A06) compared to the evaluation set, which contains different attacks and presents a generalization challenge with new types of attacks. 
We also observe differences between attacks and genuine speech across gender types, indicating that gender significantly influences the statistical variation of both attacks and genuine speech within our time embedding approach. For example, the attacks A01 and A06 occupy different areas for males and females.
These time embeddings serve as input to our CM systems. We distinguish between two types: the GD system, where separate CM systems are created for each gender, and the GI system, where only a single CM system is used.
\newline
\textbf{Training Details:}
An exhaustive grid search was conducted to identify the optimal layer sizes. 
The resulting sizes of the FC (fully connected) layers in the proposed CM system are as follows, as illustrated in \autoref{fig:Female_Male_CM}. 
The number of neurons in each layer was set to 5, 40, and 1 for layers (a), (b), and (c), respectively, for the male CM, and 10, 40\sep[]80, and 48\sep[]32 neurons for layers (d), (e), and (f), respectively, for the female\sep[]GI CM systems. 
The dropout probability was set to 0.2. 
The male CM networks were trained with a batch size of 256 for 300 epochs, while the female\sep[]GI CM systems were trained with a batch size of 32\sep128 for 100\sep200 epochs, respectively. To address the issue of data imbalance, we employed the SMOTE-SVM algorithm \cite{nguyen2011borderline} and enhanced generalization using Stochastic Gradient Descent \cite{zhou2020towards}. The optimal performance of CM models based on time domain embedding is detailed in \autoref{tab:CM_EER}. As observed in \autoref{tab:CM_EER}, the GD system exhibits a significantly superior performance compared to the GI system for the evaluation set. The CM implementation demonstrates a notable enhancement in performance compared to \cite{karo2023compact}.
\begin{table}[t]
\centering
\caption{EER performance of the CM systems, using gender recognition model\sep[]true gender labels, and gender independent.}
\label{tab:CM_EER}
\begin{tabular}{lcc}
\hline
\textbf{Subset} & \textbf{Gender System} & \textbf{EER (\%)} \\\hline
\multirow{2}{*}{Dev.} & \multirow{2}{*}{Male} & 0.55\sep[]0.55 \\
                      &      & {\scriptsize [0.21, 0.82]\sep[][0.21, 0.82]} \\ \hline
\multirow{1}{*}{Dev.} & Male \cite{karo2023compact} & 2.07 \\\hline
\multirow{2}{*}{Dev.} & \multirow{2}{*}{Female} & 0.00\sep[]0.00 \\
                      &        & {\scriptsize [0.00, 0.00]\sep[][0.00, 0.00]} \\\hline
\multirow{1}{*}{Dev.} & Female \cite{karo2023compact} & 0.12 \\\hline
\multirow{2}{*}{Dev.} & \multirow{2}{*}{GI} & 0.10 \\
                      &                 & {\scriptsize [0.00, 0.17]} \\\hline
\multirow{2}{*}{Dev.} & \multirow{2}{*}{GD} & 0.26\sep[]0.26 \\
                      &               & {\scriptsize [0.05, 0.55]\sep[][0.05, 0.55]} \\\hline
\multirow{2}{*}{Eval.} & \multirow{2}{*}{Male} & 8.52\sep[]8.67 \\
                       &      & {\scriptsize [7.10, 9.97]\sep[][7.10, 10.19]} \\\hline
\multirow{1}{*}{Eval.} & Male \cite{karo2023compact} & 12.09 \\\hline
\multirow{2}{*}{Eval.} & \multirow{2}{*}{Female} & 10.22\sep[]10.12 \\
                       &        & {\scriptsize [8.63, 12.22]\sep[][8.61, 12.03]} \\\hline
\multirow{1}{*}{Eval.} & Female \cite{karo2023compact} & 12.99 \\\hline
\multirow{2}{*}{Eval.} & \multirow{2}{*}{GI} & 10.21 \\
                       &                 & {\scriptsize [8.77, 11.92]} \\\hline
\multirow{2}{*}{Eval.} & \multirow{2}{*}{GD} & 9.68\sep[]9.68 \\
                       &               & {\scriptsize [8.50, 11.07]\sep[][9.11, 10.24]} \\
\hline
\end{tabular}
\end{table}

\subsection{Automatic Speaker Verification}
\begin{figure}[t]
    \centering
    \includegraphics[width=8.6cm]{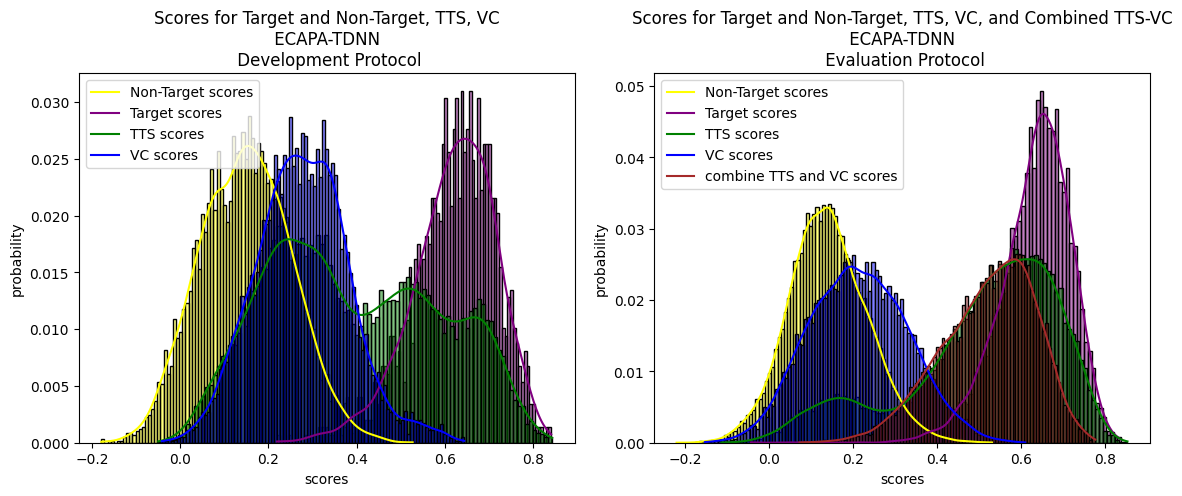}
    \caption{Histogram and \textit{kernel density estimation} (KDE) of attack analysis for ASVSpoof2019 on the ECAPA-TDNN system}
\label{fig:Attacks_ASVSpoof_ECAPA_TDNN_ALL}
\end{figure}

As previously noted, the ECAPA-TDNN system is utilized as the ASV system for each gender.
The minimum DCF in \autoref{tab:ASV_Prformance} for our ASV systems was calculated with $\mathrm{\pi_{tar}}= 0.99, \mathrm{\pi_{non}} = 0.01$, and without the influence of spoofing attacks (i.e., $\mathrm{\pi_{spoof}} = 0$, meaning genuine speech only). 
The minimum DCF is presented in \autoref{tab:ASV_Prformance}, while the score PMFs for each class, target, non-target zero-effort, and different types of spoofing attacks are illustrated in \autoref{fig:Attacks_ASVSpoof_ECAPA_TDNN_ALL}.
\begin{table}[t]
    \centering
     \caption{ECAPA-TDNN system performance on Asvspoof2019}
    \begin{tabular}{ccccc}
        \hline
        \textbf{Subset} & \textbf{Gender} & \textbf{minDCF} & \textbf{EER (\%)} & \textbf{EER Threshold}\\
        \hline
        \multirow{2}{*}{Dev.} & \multirow{2}{*}{Males}  & 0.002 & 0.34 & 0.261 \\ &
         & {\scriptsize [0.000,0.003]} &  {\scriptsize [0.0, 0.95]} &  {\scriptsize [0.230, 0.284]} \\
        \multirow{2}{*}{Dev.} & \multirow{2}{*}{Females} & 0.008 & 1.08 & 0.397 \\ &
         & {\scriptsize [0.000,0.011]} &  {\scriptsize [0.00,2.39]} &  {\scriptsize [0.334,0.459]} \\
        \multirow{2}{*}{Dev.} & \multirow{2}{*}{Combined}  & 0.009 & 1.35 & 0.375\\ &
        & {\scriptsize [0.000,0.014]} &  {\scriptsize [0.35,2.55]} &  {\scriptsize [0.330,0.411]} \\
        \multirow{2}{*}{Eval.} & \multirow{2}{*}{Males}  & 0.003 & 0.62 & 0.352\\ &
        & {\scriptsize [0.000,0.005]} &  {\scriptsize [0.16,1.20]} &  {\scriptsize [0.327,0.383]} \\
        \multirow{2}{*}{Eval.} & \multirow{2}{*}{Females} & 0.005 & 0.78 & 0.401 \\ &
         & {\scriptsize [0.001,0.007]} &  {\scriptsize [0.40,1.32]} &  {\scriptsize [0.372,0.431]} \\
        \multirow{2}{*}{Eval.} & \multirow{2}{*}{Combined}   & 0.005 & 0.80 & 0.384 \\ &
         & {\scriptsize [0.002,0.006]} &  {\scriptsize [0.43,1.21]} &  {\scriptsize [0.364,0.407]} \\
        \hline
    \end{tabular}
    \label{tab:ASV_Prformance}
\end{table}
From~\autoref{fig:Attacks_ASVSpoof_ECAPA_TDNN_ALL} it is evident that the ASV system is more vulnerable to TTS and TTS-VC combination attacks. These observations corroborate the findings presented in \cite{jung24d_interspeech}, wherein ASV-based DNN systems demonstrated heightened vulnerability to TTS attacks, while GMM-based systems exhibited heightened vulnerability to VC attacks.
As illustrated in \autoref{tab:ASV_Prformance}, the optimal performance is attained when the system is deployed with distinct thresholds for the male and female genders. In the case of gender-independence, the threshold is situated between the male and female thresholds. In both the development and evaluation sets, the threshold for the female gender is higher than that for the male gender. In order to implement an ASV system with distinct thresholds for males and females, it is necessary to ascertain the gender of the input speech. This may be accomplished in two ways: first, by assuming that the gender labels are known; second, by applying a gender recognition system (see \autoref{subsec:GenderRecognition}).
\begin{table}[t]
  \caption{ECAPA-TDNN performance on the ASVspoof2019 LA database in terms of minDCF and EER. The threshold is determined based on the EER of each database subset,
  using gender classification model\sep[]true gender labels}
  \label{tab:minDCF_EER_ECAPA_TDNN}
  \centering
\begin{tabular}{m{2em}m{3.3em}cc}
\hline
\textbf{Subset} & \textbf{System} & \textbf{EER(\%)} & \textbf{minDCF} \\\hline
\multirow{2}{*}{Dev.} & \multirow{2}{*}{GD} & 0.90\sep0.90 & 0.006\sep0.006 \\
                      &                           & {\scriptsize [0.15, 1.96]\sep[][0.15, 1.96]} & {\scriptsize [0.001, 0.013]\sep[][0.001, 0.013]} \\\hline
\multirow{2}{*}{Dev.} & \multirow{2}{*}{GI} & 1.34 & 0.009 \\
                      &                           & {\scriptsize [0.39, 2.48]} & {\scriptsize [0.003, 0.016]} \\\hline
\multirow{2}{*}{Eval.} & \multirow{2}{*}{GD} & 0.73\sep0.73 & 0.004\sep0.004 \\
                       &                           & {\scriptsize [0.41, 1.08]\sep[][0.42, 1.09]} & {\scriptsize [0.002, 0.007]\sep[][0.002, 0.007]} \\\hline
\multirow{2}{*}{Eval.} & \multirow{2}{*}{GI} & 0.80 & 0.005 \\
                       &                           & {\scriptsize [0.47, 1.18]} & {\scriptsize [0.002, 0.007]} \\
\hline
\end{tabular}
\end{table}
The impact of gender recognition on ASV system performance was examined by comparing GD and GI systems. The CM threshold was determined for each set in the database. Similarly to the ASV case, for the CM GI scenario, a single threshold was applied to the entire dataset, comprising both male and female subjects. In contrast, in the CM GD scenario, two thresholds were utilized, one for each gender. The results based on the ASVspoof2019 LA database, with separate analyzes for male and female categories, are presented in \autoref{tab:minDCF_EER_ECAPA_TDNN}.

\subsection{Spoofing-Robust Automatic Speaker Verification}
The normalized minimum t-DCF ASV-constrained results are detailed for the GD CM in Tables \ref{tab:min_tdcf_male},~\ref{tab:min_tdcf_female} and for the GI CM in Table~\ref{tab:min_tdcf_both}. The results are for the proposed gender recognition model and for true gender labels (perfect gender recognition assumption). {The results demonstrate that the GD system architecture outperforms the GI system for the evaluation set, aligning with findings in \cite{karo2023compact}.} Furthermore, there is no degradation in terms of min t-DCF for the development set when using a gender recognition model. For the evaluation set, the degradation is minor.
\begin{table}[t]
\caption{Normalized asv-constrained $\mathrm{\min\ \tDCF}$ on the spoofing-robust speaker verification system, using gender recognition model\sep[]true gender labels and gender independent system for males.}
\label{tab:min_tdcf_male}
\centering
\begin{tabular}{llcl}
\hline
\textbf{Subset} & \textbf{System} & \multicolumn{2}{c}{\textbf{Min norm t-DCF}} \\ 
\hline
\multirow{2}{*}{Dev.} & \multirow{2}{*}{GD} & \multicolumn{2}{c}{0.0105\sep0.0105} \\
                      &                             & \multicolumn{2}{c}{{\scriptsize [0.0023, 0.0160]\sep[][0.0023, 0.0160]}} \\\hline
\multirow{2}{*}{Dev.} & \multirow{2}{*}{GI} & \multicolumn{2}{c}{0.0012} \\
                      &                               & \multicolumn{2}{c}{{\scriptsize [0.0000, 0.0014]}} \\\hline
\multirow{2}{*}{Eval.} & \multirow{2}{*}{GD} & \multicolumn{2}{c}{0.2200\sep0.2185} \\
                       &                             & \multicolumn{2}{c}{{\scriptsize [0.1881, 0.2439]\sep[][0.1871, 0.2413]}} \\\hline
\multirow{2}{*}{Eval.} & \multirow{2}{*}{GI} & \multicolumn{2}{c}{0.3223} \\
                       &                               & \multicolumn{2}{c}{{\scriptsize [0.2892, 0.3357]}} \\
\hline
\end{tabular}
\end{table}
\begin{table}[t]
\caption{Normalized asv-constrained $\mathrm{\min \tDCF}$ on the spoofing-robust speaker verification system, using gender recognition model\sep[]true gender labels and gender independent system for females.}
\label{tab:min_tdcf_female}
\centering
\begin{tabular}{llcl}
\hline
\textbf{Subset} & \textbf{System} & \multicolumn{2}{c}{\textbf{Min norm t-DCF}} \\ \hline
\multirow{2}{*}{Dev.} & \multirow{2}{*}{GD} & \multicolumn{2}{c}{0.0000\sep0.0000} \\
                      &                             & \multicolumn{2}{c}{{\scriptsize [0.0000, 0.0000]\sep[][0.0000, 0.0000]}} \\\hline
\multirow{2}{*}{Dev.} & \multirow{2}{*}{GI} & \multicolumn{2}{c}{0.0018} \\
                      &                               & \multicolumn{2}{c}{{\scriptsize [0.0000, 0.0023]}} \\\hline
\multirow{2}{*}{Eval.} & \multirow{2}{*}{GD} & \multicolumn{2}{c}{0.2949\sep0.2927} \\
                       &                             & \multicolumn{2}{c}{{\scriptsize [0.2683, 0.3171]\sep[][0.2656, 0.3158]}} \\\hline
\multirow{2}{*}{Eval.} & \multirow{2}{*}{GI} & \multicolumn{2}{c}{0.3123} \\
                       &                               & \multicolumn{2}{c}{{\scriptsize [0.2902, 0.3207]}} \\
\hline
\end{tabular}
\end{table}

\begin{figure}[t]
\centering
\includegraphics[width=8.6cm]{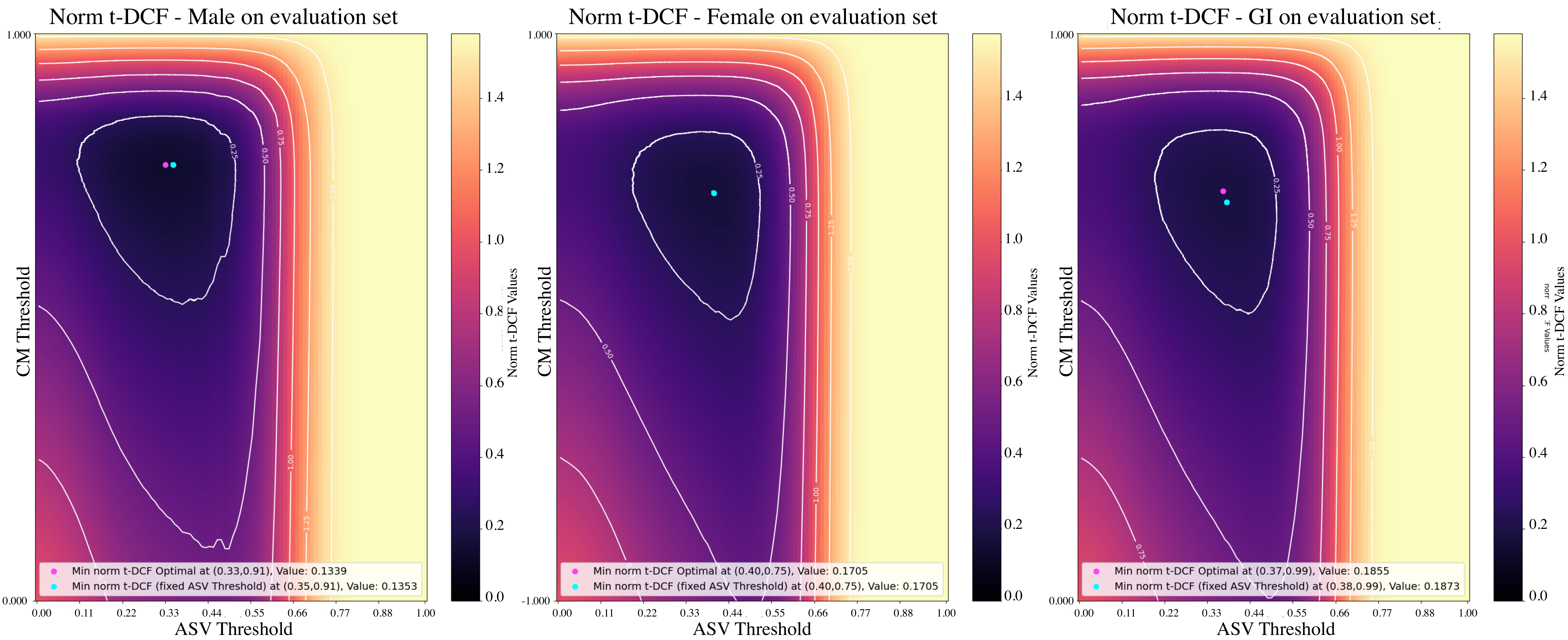}
  \caption{The minimum normalized unconstrained t-DCF
: left - males, middle - females, and right for GI CM system.}
\label{fig:all_eval_set_togther}
\end{figure}

\begin{table}[t]
\centering
\caption{Normalized asv-constrained  $\mathrm{\min \tDCF}$ on the spoofing-robust speaker verification system, using gender classification model\sep[]true gender labels and gender independent system for both genders}
\label{tab:min_tdcf_both}
\begin{tabular}{llcl}
\hline
\textbf{Subset} & \textbf{System} & \multicolumn{2}{c}{\textbf{Norm min t-DCF}} \\ \hline
\multirow{2}{*}{Dev.} & \multirow{2}{*}{GD} & \multicolumn{2}{c}{0.0039\sep0.0039} \\
                      &                             & \multicolumn{2}{c}{{\scriptsize [0.0005, 0.0124]\sep[][0.0005, 0.0124]}} \\\hline
\multirow{2}{*}{Dev.} & \multirow{2}{*}{GI} & \multicolumn{2}{c}{0.0016} \\
                      &                               & \multicolumn{2}{c}{{\scriptsize [0.0000, 0.0022]}} \\\hline
\multirow{2}{*}{Eval.} & \multirow{2}{*}{GD} & \multicolumn{2}{c}{0.2709\sep0.2693} \\
                       &                             & \multicolumn{2}{c}{{\scriptsize [0.2499, 0.2923]\sep[][0.2484, 0.2914]}} \\\hline
\multirow{2}{*}{Eval.} & \multirow{2}{*}{GI} & \multicolumn{2}{c}{0.3178} \\
                       &                               & \multicolumn{2}{c}{{\scriptsize [0.2990, 0.3259]}} \\
\hline
\end{tabular}
\end{table}
The analysis of results across various thresholds for the ASV and the implemented CM systems using the normalized unconstrained t-DCF metric is consistent with the methodology proposed in \cite{kinnunen2020tandem}. \autoref{fig:all_eval_set_togther} illustrates the impact of these thresholds on the normalized unconstrained min t-DCF for each SASV system. It is evident that the EER operating point (cyan) is not the optimal point, and a better operating point can be attained (red), except in the female plot, where the cyan and red points coincide. It should be noted that a non-linear y-axis has been employed to provide a higher resolution for the pertinent values, based on the software package published in \cite{ASV_Soft}. The normalized min a-DCF for each system is presented in~\autoref{tab:a_dcf}. The fixed CM threshold is pre-set according to the optimal threshold for each subset.

\begin{table}[h!]
\centering
\caption{Normalized $\mathrm{\min \aDCF}$ on the spoofing-robust speaker verification system, using gender classification model\sep[]true gender labels and gender independent system for both gender}
\label{tab:a_dcf}
\begin{tabular}{lccc}
\hline
\textbf{Subset} & \textbf{System} & \multicolumn{2}{c}{\textbf{Norm min a-DCF}} \\ \hline
\multirow{2}{*}{Dev.} & \multirow{2}{*}{GD} & \multicolumn{2}{c}{0.0064\sep0.0064} \\
                      &                     & \multicolumn{2}{c}{\scriptsize [0.0102,0.0138]} \\
\multirow{2}{*}{Dev.} & \multirow{2}{*}{GI} & \multicolumn{2}{c}{0.0046}         \\
                      &                     & \multicolumn{2}{c}{\scriptsize [0.0000,0.0074]} \\
\multirow{2}{*}{Eval.} & \multirow{2}{*}{GD} & \multicolumn{2}{c}{0.1250\sep0.1216} \\
                      &                     & \multicolumn{2}{c}{\scriptsize [0.0923, 0.1679]\sep[][0.0893, 0.1659]} \\
\multirow{2}{*}{Eval.} & \multirow{2}{*}{GI} & \multicolumn{2}{c}{0.1684}         \\
                      &                     & \multicolumn{2}{c}{\scriptsize [0.1437, 0.1996]} \\
\hline
\end{tabular}
\end{table}

\subsection{Fusion Countermeasures}
The application of fusion models is becoming more prevalent in enhancing spoofing detection in the development of CM systems for SASV tasks. These models provide a more comprehensive and accurate detection mechanism by combining multiple subsystems, each of which may employ different feature extraction techniques (such as our PMF-based time embedding approach) or different classifiers. The subsystems may be combined at various stages, including the scoring level or the feature stage. The traditional LFCC features have been shown to perform well in the ASVSpoof2019 challenge. Three frequency-based systems, based on the ResNet-18 architecture with temporal attention, were adopted from \cite{he2016deep}. The LFCC features serve as the system input, and three optimization losses, namely OCS, AMS, and cross-entropy, were applied. The objective was to ascertain whether the overall performance is enhanced when these systems are combined with the proposed GD CM system, which is based on a time embedding approach. The performance of the individual systems is detailed in \autoref{tab:CM_EER_FREQ}. In the score-level fusion, the scores can be combined using a variety of methods, including weighted averaging or inputting the scores into a new classifier. The range of scores produced by the chosen loss optimization function may fall within either the interval $[0, 1]$ or the interval $[-1, 1]$. To combine these scores for fusion, the $s_{new}=(s_{old}+1)/2$ transformation is applied, whereby the values from the interval $[-1, 1]$ are mapped to the interval $[0, 1]$.
\begin{table}[t]
\centering
\caption{Comparison of EERs across ResNet-18 with temporal attention LFCC-based systems from \cite{zhang2021one} and our gender-dependent system for males\sep[]females\sep[]both genders.}
\begin{tabular}{lcc}
\hline
\textbf{Subset} & \textbf{System} & \textbf{EER (\%)} \\ \hline
\multirow{2}{*}{Dev.} & \multirow{2}{*}{GD} & 0.55\sep0.00\sep0.10 \\
                      &                       & {\scriptsize [0.21, 0.82] / [0.00, 0.00] / [0.05, 0.55]} \\\hline
\multirow{2}{*}{Dev.} & \multirow{2}{*}{Softmax} & 0.34\sep0.05\sep0.15 \\
                      &                          & {\scriptsize [0.00, 0.89] / [0.00, 0.26] / [0.00, 0.39]} \\\hline
\multirow{2}{*}{Dev.} & \multirow{2}{*}{AMSoftmax} & 0.74\sep0.23\sep0.40 \\
                      &                            & {\scriptsize [0.13, 1.54] / [0.05, 0.54] / [0.10, 0.77]} \\\hline
\multirow{2}{*}{Dev.} & \multirow{2}{*}{OCSoftmax} & 0.54\sep0.17\sep0.29 \\
                      &                             & {\scriptsize [0.07, 1.39] / [0.00, 0.54] / [0.05, 0.67]} \\\hline
\multirow{2}{*}{Eval.} & \multirow{2}{*}{GD} & 8.67\sep10.12\sep9.68 \\
                       &                        & {\scriptsize [7.10, 10.19] / [8.61, 12.03] / [9.11, 10.24]} \\\hline
\multirow{2}{*}{Eval.} & \multirow{2}{*}{Softmax} & 4.47\sep5.28\sep5.03 \\
                       &                           & {\scriptsize [3.80, 5.22] / [4.38, 6.24] / [4.41, 5.78]} \\\hline
\multirow{2}{*}{Eval.} & \multirow{2}{*}{AMSoftmax} & 3.32\sep3.09\sep3.16 \\
                       &                             & {\scriptsize [2.59, 4.09] / [2.39, 3.75] / [2.67, 3.70]} \\\hline
\multirow{2}{*}{Eval.} & \multirow{2}{*}{OCSoftmax} & 2.40\sep2.05\sep2.18 \\
                       &                              & {\scriptsize [1.90, 2.91] / [1.47, 2.60] / [1.74, 2.54]} \\
\hline 
\end{tabular}
\label{tab:CM_EER_FREQ}
\end{table}

\begin{table*}[t]
\centering
\caption{Comparison of EERs between the weighted average scoring fusion based on the GD CM system and the systems from \cite{zhang2021one}, for males, females, and both genders.}
\begin{tabular}{lccccc}
\hline
\textbf{System}  & \makecell{\textbf{Dev. Set} \\ \textbf{EER (\%)}} & \makecell{\textbf{Eval Set} \\ \textbf{EER (\%)}  ($\alpha_{Dev}$)} & \makecell{\textbf{Eval Set} \\ \textbf{EER (\%)}  ($\alpha_{Eval}$)} & \textbf{$\alpha_{Dev}$} & \textbf{$\alpha_{Eval}$}\\ \hline
Softmax+GD   & 0.00\sep0.00\sep0.03  & 4.71\sep5.83\sep5.30 &  4.25\sep5.18\sep4.90 & 0.36\sep0.58\sep0.39   & 0.03\sep0.05\sep0.05  \\ 
AMSoftmax+GD  & 0.16\sep0.00\sep0.06 & 4.24\sep4.65\sep4.57 & 3.10\sep3.08\sep3.09  & 0.64\sep0.56\sep0.61 & 0.01\sep0.01\sep0.01  \\ 
OCSoftmax+GD   &  0.12\sep0.00\sep0.06 & 3.53\sep3.76\sep4.06 &  2.26\sep2.02\sep2.10 &  0.75\sep0.64\sep0.83 & 0.03\sep0.01\sep0.01 \\
\hline
\end{tabular}
\label{tab:CM_EER_WEIGTHED_SCORE}
\end{table*}

\subsubsection*{Fusion - Weighted Average Scoring}
The weighted average is a suitable method for performing score fusion. The fused score is computed using the following expression:
\begin{equation}
   \mathrm{S} = \alpha \cdot \mathrm{S}_{\mathrm{GD}} + (1 - \alpha) \cdot \mathrm{S}_{\mathrm{LFCC}} 
   \label{eq:Weighted_Average_Scoring}
\end{equation}
$\alpha$ is a hyper-parameter ranging between 0 and 1, with the optimal value determined based on the performance of the development set. Nevertheless, due to potential discrepancies in distribution between the development and evaluation sets, selecting $\alpha$ based on the former may not yield optimal performance on the latter. Consequently, the minimum EER was also evaluated by estimating the optimal $\alpha$ value on the evaluation set. 
The results presented in \autoref{tab:CM_EER_WEIGTHED_SCORE} for choosing the optimal $\alpha$ value are shown separately for males, females, and both genders.
It can be seen that the optimal $\alpha$ value, selected based on the evaluation set, resulted in enhanced performance. However, selecting $\alpha$ from the development set yields degradation in performance on the evaluation set due to the discrepancy between the attacks in the development and evaluation sets. 
In case multiple optimal $\alpha$ values were identified for the development set, the minimum value is chosen.
Moreover, the lowest  $\alpha$ value found for the evaluation set attenuates the influence of the scores of the males, females, and both genders systems, which are based on time embeddings. This suggests that CM systems lack supplementary information when a weighted score fusion method is applied.
\subsubsection*{Fusion - Classifier based Scoring}
\begin{figure}[t]
    \centering
    \includegraphics[width=8.6cm]{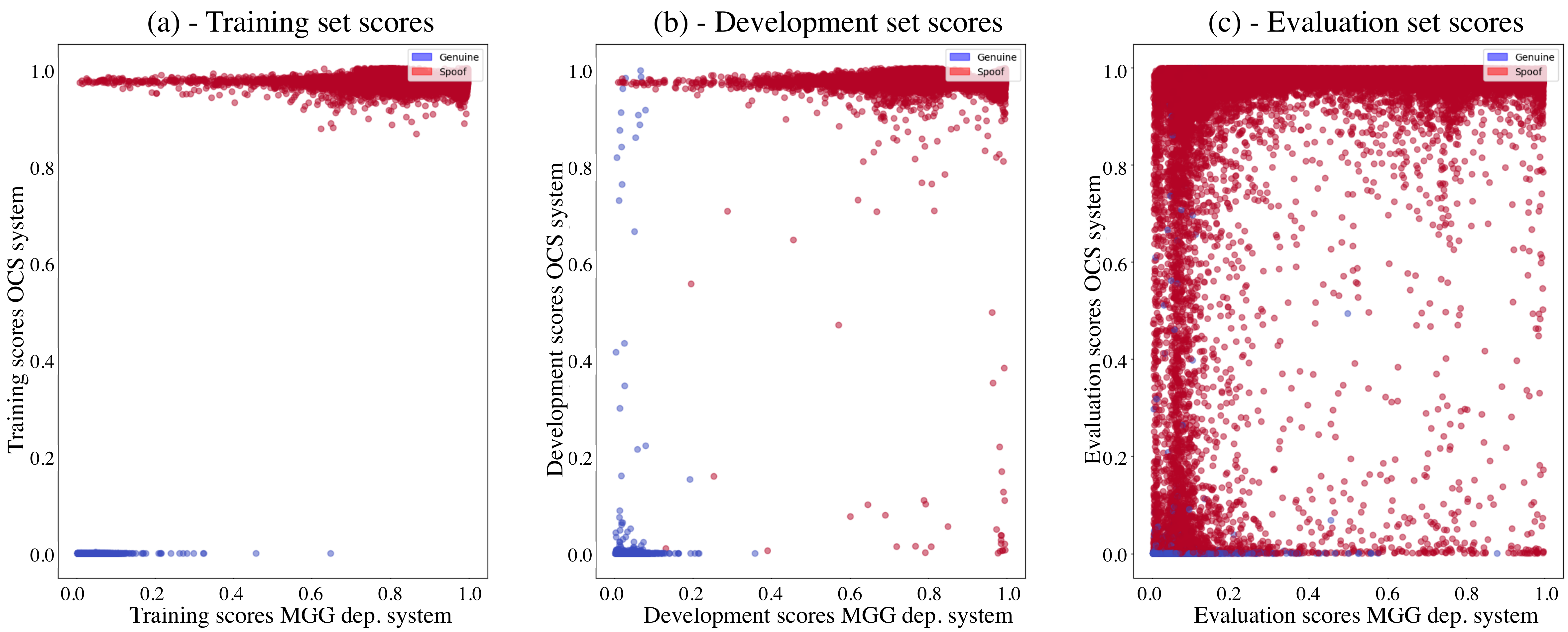}
    \caption{Scores of GD CM System vs. OCSoftmax CM System.}
    \label{fig:scores_visualization}
\end{figure}

In comparison to the preceding weighted average score fusion methodology, this approach integrates the assessments from the two CM systems through the use of a classifier, thereby generating a novel fused score. The classifiers selected for score fusion include LR, kNN, RF, SVM, balanced RF, XGB, and RUSBoost. In each configuration, the classifier utilized the scores from each CM system as an input. \autoref{fig:scores_visualization} illustrates the visualization scores from the OCSoftmax CM system and the  GD CM system.
It can be seen that the training set (a) and development set (b) exhibit a clear separation due to their compatibility with the attack types. The evaluation set (c) displays less distinct separation. The GD scores exhibit greater separation within the development set in comparison to the OCSoftmax scores. Furthermore, \autoref{fig:scores_visualization} (c) illustrates the significance of hyper-parameter selection in the classifier that generates the fused score, as it can considerably influence the capability to generalize and detect novel attacks in the evaluation set. The hyper-parameters for each classifier candidate were optimized through the use of a grid search method, thereby achieving optimal performance on the development set. Once the optimal configuration has been identified, the top-performing classifier is subjected to further examination of the evaluation set. To investigate the discrepancy between the training and development sets in comparison to the evaluation set, and to examine the complementary information captured by the time embeddings approach to LFCC-based features, separate grid searches were conducted: one on the development set (\autoref{tab:CM_EER_fUSION_dev}) and another on the evaluation set (\autoref{tab:CM_EER_fUSION_eval}). The grid search tuning on the evaluation set demonstrates the discrepancy between the conditions in the development and evaluation sets and the challenge of generalization.
\begin{table}[t]
\caption{Comparison of EERs between score fusion based on the GD CM system and the systems from \cite{zhang2021one} using machine learning classifiers for males, females, and both genders, tuning validated on the dev. set.}
\centering
\begin{tabular}{lcc}
\hline
\textbf{Subset} & \textbf{System} &   \textbf{EER (\%)} \\ \hline
\multirow{2}{*}{Dev.} & \multirow{2}{*}{Softmax+GD} & 0.08\sep0.00\sep0.02\\
                       &                                        & {\scriptsize [0.00, 0.24] / [0.00, 0.00] / [0.00, 0.07]} \\\hline
\multirow{2}{*}{Dev.} & \multirow{2}{*}{AMSoftmax+GD} & 0.01\sep0.00\sep0.00\\
                       &                                        & {\scriptsize [0.00, 0.06] / [0.00, 0.00] / [0.00, 0.02]} \\\hline
\multirow{2}{*}{Dev.} & \multirow{2}{*}{OCSoftmax+GD} &  0.09\sep0.00\sep0.03\\
                       &                                        & {\scriptsize [0.00, 0.21] / [0.00, 0.00] / [0, 0.06]} \\\hline
\multirow{2}{*}{Eval.} & \multirow{2}{*}{Softmax+GD} & 4.70\sep5.54\sep5.28\\
                       &                                        & {\scriptsize [4.00, 5.46] / [5.06, 6.32] / [4.86, 5.89]} \\\hline
\multirow{2}{*}{Eval.} & \multirow{2}{*}{AMSoftmax+GD} &  3.48\sep3.45\sep3.46\\
                       &                                        & {\scriptsize [2.84, 4.58] / [2.83, 4.21] / [2.97, 4.06]} \\\hline
\multirow{2}{*}{Eval.} & \multirow{2}{*}{OCSoftmax+GD} &  2.62\sep2.61\sep2.61 \\
                       &                                        & {\scriptsize [1.95, 3.63] / [2.32, 3.20] / [2.31, 3.17]} \\
\hline
\end{tabular}
\label{tab:CM_EER_fUSION_dev}
\end{table}
\begin{table}[t]
\caption{Comparison of EERs between score fusion based on the GD CM system and the systems from \cite{zhang2021one} using machine learning classifiers for males, females, and both genders, tuning validated on the eval. set.}
\centering
\begin{tabular}{lcc}
\hline
\textbf{Subset} & \textbf{System} &   \textbf{EER (\%)} \\ \hline
\multirow{2}{*}{Dev.} & \multirow{2}{*}{Softmax+GD} & 0.15\sep0.01\sep0.06 \\
                      &                                          & {\scriptsize [0.02,0.32] / [0.00,0.02] / [0.00, 0.12]} \\\hline
\multirow{2}{*}{Dev.} & \multirow{2}{*}{AMSoftmax+GD} & 0.20\sep0.11\sep0.14 \\
                      &                                           & {\scriptsize [0.00, 0.40] / [0.00, 0.48] / [0.00, 0.39]} \\\hline
\multirow{2}{*}{Dev.} & \multirow{2}{*}{OCSoftmax+GD} & 0.34\sep0.59\sep0.50 \\
                      &                                           & {\scriptsize [0.01, 0.67] / [0.11, 1.33] / [0.14, 1.06]} \\\hline
\multirow{2}{*}{Eval.} & \multirow{2}{*}{Softmax+GD} & 4.11\sep4.64\sep4.47\\
                       &                                        & {\scriptsize [3.40, 5.02] / [3.94, 5.30] / [3.89, 5.04]} \\\hline
\multirow{2}{*}{Eval.} & \multirow{2}{*}{AMSoftmax+GD} & 2.97\sep2.91\sep2.96\\
                       &                                        & {\scriptsize [2.30, 3.61] / [2.30, 3.57] / [2.40, 3.35]} \\\hline
\multirow{2}{*}{Eval.} & \multirow{2}{*}{OCSoftmax+GD} & 2.01\sep1.78\sep1.85\\
                       &                                        & {\scriptsize [1.52, 2.40] / [1.40, 2.27] / [1.54, 2.21]} \\
\hline
\end{tabular}
\label{tab:CM_EER_fUSION_eval}
\end{table}
As demonstrated in Tables \ref{tab:CM_EER_fUSION_dev} and \ref{tab:CM_EER_fUSION_eval}, the fusion system exhibits superior performance compared to both the LFCC-based and time embeddings-based systems on the development and evaluation sets when the optimal hyper-parameters are identified using those respective sets. However, when hyper-parameters were selected based on the development set, the evaluation set results were inferior to those of the LFCC-based systems. This can be attributed to the fact that all classifiers achieve near-perfect performance on the training and development sets, thereby making it challenging to identify the optimal model parameters during grid search tuning with the development set.

\section{Conclusions and future work}
\label{sec:Conclusions_and_future_work}
This paper presents a novel, comprehensive, spoofing-robust automatic speaker verification system. Our results demonstrate the efficacy of time-domain embeddings for anti-spoofing tasks and illustrate their potential for application in other domains, such as gender recognition. Our results demonstrate that the effective utilization of time embeddings in the training set facilitates precise differentiation between male and female speakers. Furthermore, our results indicate that GD systems exhibit enhanced robustness compared to GI systems when employing time-domain embedding techniques. In matched conditions, the time-embedding-based CM system exhibits near-perfect performance. However, its generalization capabilities are significantly degraded when the conditions are not matched. By combining the proposed time embedding approach with widely used frequency-based CM systems to leverage both the frequency and time domains, we demonstrate that time-domain embeddings do not provide additional information when using linear classifiers such as weighted average scoring. We also illustrate the sensitivity involved in hyper-parameter selection during the fusion stage. The fusion of the two systems yielded enhanced overall results on the evaluation set when the hyper-parameters were optimized on the evaluation set. However, a notable degree of overlap in the confidence interval range persists compared to the LFCC-based CM system.
To the best of our knowledge, this paper represents the first effort to investigate the integration of time embeddings through fusion and to design a SASV system utilizing time embeddings. This research introduces a novel approach to time embeddings in a SASV system. Further investigation is warranted with respect to system architecture, the selection of similarity measures, and the application of filters to the construction of embeddings. While previous works have primarily focused on the embeddings themselves, this study presents an entire system. As demonstrated in \cite{karo24_odyssey}, the embeddings are explainable, providing an essential foundation for explainable SASV. In the future, we aim to investigate alternative SASV configurations, both tandem and non-tandem, as well as novel fusion strategies to enhance the generalization capabilities of the SASV system.
\section{Acknowledgments}
This work is supported by the Israel Innovation Authority under project numbers 82457 and 82458.

\ifCLASSOPTIONcaptionsoff
  \newpage
\fi
\bibliographystyle{IEEEtran} 
\bibliography{reference}
\begin{IEEEbiography}[{\includegraphics[width=1in,height=1.25in,clip,keepaspectratio]{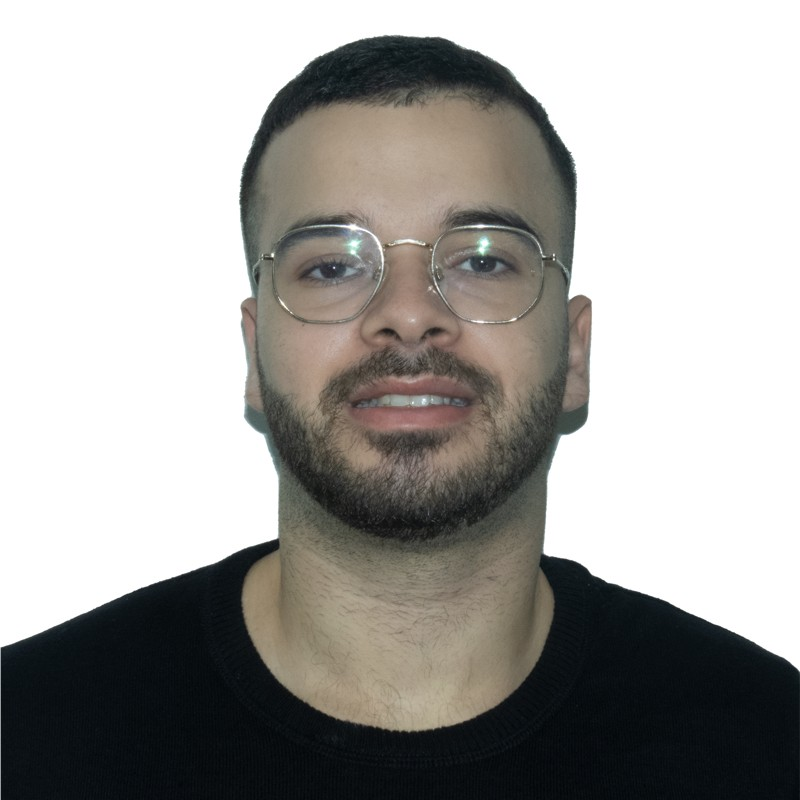}}]{Avishai Weizman}
Avishai Weizman received his B.Sc. in Electrical Engineering from Afeka Tel-Aviv Academic College of Engineering, Israel, in 2019, and these days he is studying for an M.Sc. in Electrical and Computer Engineering at Ben-Gurion University, Beer-Sheva, Israel. He is currently a Speech and Computer Vision Researcher and Team Lead. His research interests include speech anti-spoofing, speaker verification, speech synthesis, speech-to-text, object detection, image classification, and data analysis using AI and machine learning models.
\end{IEEEbiography}
\vfill
\begin{IEEEbiography}[{\includegraphics[width=1in,height=1.25in,clip,keepaspectratio]{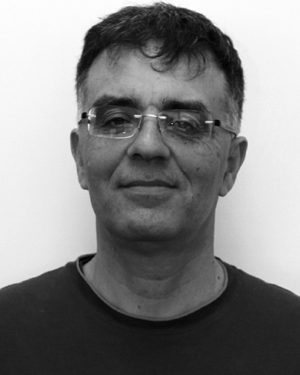}}]{Yehuda Ben-Shimol}
Yehuda Ben-Shimol (Member, IEEE) received the B.Sc, the M.Sc, and the Ph.D. degrees (Hons.) from Ben-Gurion University of the Negev, Beer-Sheva, Israel, all in electrical and computer engineering. He is currently a Senior Lecturer at the School of Electrical and Computer Engineering, Ben-Gurion University of the Negev. His main areas of interest are design and analysis and performance evaluation of communication networks, computer architecture, machine and deep learning (centralized and distributed) and neuromorphic computing.
\end{IEEEbiography}
\vfill
\begin{IEEEbiography}[{\includegraphics[width=1in,height=1.25in,clip,keepaspectratio]{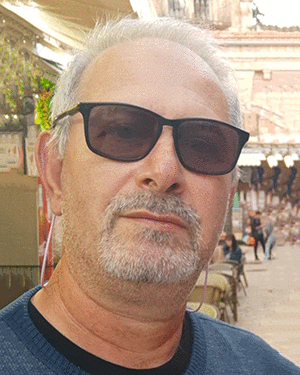}}]{Itshak Lapidot}
Itshak Lapidot received the B.Sc., M.Sc., and Ph.D. degrees from the Electrical and Computer Engineering Department, Ben-Gurion University, Beer-Sheva, Israel. He held a postdoctoral position with IDIAP Switzerland. He was a Lecturer with the Electrical and Electronics Engineering Department, Sami Shamoon College of Engineering, Beer-Sheva, Israel. He was a Researcher with the Laboratoire Informatique d’Avignon (LIA), University of Avignon, Avignon, France. He is a Member of the Electrical Engineering Department, Afeka, Tel-Aviv Academic College of Engineering, Israel, and a Researcher with the Afeka Center of Language Processing. He is also an associated Researcher with LIA, Avignon. His research interests include speaker diarization, speaker clustering, speaker verification, anti-spoofing, and data assessment.
\end{IEEEbiography}
\vfill
\end{document}